\documentclass[twocolumn,aps,prb,showpacs,preprintnumbers,amsmath,amssymb]{revtex4}

\usepackage{epsfig}
\usepackage{graphicx}% Include figure files
\usepackage{dcolumn}% Align table columns on decimal point
\usepackage{bm}% bold math

\newcommand{\sw}{\sqrt{\omega}}

\renewcommand{\k}{{\bf k}}
\newcommand{\kNull}{{\bf k}_0}
\newcommand{\q}{{\bf q}}

\newcommand{\tb}{\tilde b}
\newcommand{\tq}{\tilde q}

\def\gl{\lower.35em\hbox{$\stackrel{\textstyle>}{\textstyle<}$}}

\def\gapp{\lower.35em\hbox{$\stackrel{\textstyle>}{\sim}$}} 
\def\lapp{\lower.35em\hbox{$\stackrel{\textstyle<}{\sim}$}} 

\begin{document}
\title{Fermi Liquid Theory of a Fermi Ring}
\author{T.~Stauber$^1$, N.~M.~R. Peres$^2$, F.~Guinea$^{1}$, and A.~H.~Castro Neto$^{3,4}$}

\affiliation{$^1$Instituto de Ciencia de Materiales de
Madrid. CSIC. Cantoblanco. E-28049 Madrid, Spain}

\affiliation{$^2$Center of Physics and Departamento de
F\'{\i}sica, Universidade do Minho, P-4710-057, Braga, Portugal}

\affiliation{$^3$ Department of Physics, Harvard University, Cambridge, MA
  02138, USA}

\affiliation{$^4$Department of Physics, Boston University, 590 
Commonwealth Avenue, Boston, MA 02215,USA}

%\draft

\date{\today}
%%%%%%%%%%%%%%%%%%%%%%%%%%%%%%%%%%%%%%%%%%%%%%%%%%%%%%%%%%%%%%%%%%%%%%%%%%%%%
\begin{abstract}
We study the effect of electron-electron interactions in
the electronic properties of a biased graphene bilayer. This system 
is a semiconductor with conduction and valence bands characterized 
by an unusual ``mexican-hat'' dispersion.  
We focus on the metallic regime where the chemical potential lies in
the ``mexican-hat'' in the conduction band, leading to a topologically
non-trivial Fermi surface in the shape of a ring. 
We show that due to the unusual topology of the Fermi surface
electron-electron interactions are greatly enhanced. We show that the 
ferromagnetic instability can occur provided a low density of carriers.
We compute the electronic polarization function in the random
phase approximation and show that, while at low energies the
system behaves as a Fermi liquid (albeit with peculiar Friedel
oscillations), at high frequencies it shows
a highly anomalous response when compare to ordinary metals.
\end{abstract}
%%%%%%%%%%%%%%%%%%%%%%%%%%%%%%%%%%%%%%%%%%%%%%%%%%%%%%%%%%%%%%%%%%%%%%%%%%%%%
%
%
%
%%%%%%%%%%%%%%%%%%%%%%%%%%%%%%%%%%%%%%%%%%%%%%%%%%%%%%%%%%%%%%%%%%%%%%%%%%%%%
%%%%%%%%%%%%%%%%%%%%%%%%%%%%%%%%%%%%%%%%%%%%%%%%%%%%%%%%%%%%%%%%%%%%%%%%%%%%%
%%%%%%%%%%%%%%%%%%%%%%%%%%%%%%%%%%%%%%%%%%%%%%%%%%%%%%%%%%%%%%%%%%%%%%%%%%%%%
%%%%%%%%%%%%%%%%%%%%%%%%%%%%%%%%%%%%%%%%%%%%%%%%%%%%%%%%%%%%%%%%%%%%%%%%%%%%%
%
\maketitle
\section{Introduction}

The discovery of graphene \cite{Exp}, and the realization that it
presents very unusual electronic properties \cite{Peres06}, has
generated an enormous amount of interest in the condensed matter
community. The possibility of creating electronic devices consisting
of only a few atomic layers can open doors for a carbon-based
micro-electronics.  In this context, bilayer systems\cite{McCannFalko}
play a distinguished role for being the minimal system to produce a
gap in the electronic spectrum due to an electric field effect
\cite{McCann,GNP06}.  Recent experiments show that the electronic gap
and the chemical potential can be tuned independently of each other
and the band structure can be well described by a tight-binding model
corrected by charging effects \cite{eduardo}. The electronic gap in
these systems has been observed recently in angle resolved
photoemission (ARPES) \cite{Ohta} in epitaxially grown graphene films
on SiC crystal surfaces \cite{walt}.

Electron-electron interactions are usually neglected in single layer
graphene since they do not seem to play an important role in the
transport measurements in low magnetic fields, e. g., in the
measurements of the anomalous quantum Hall effect \cite{qhe}.  At high
magnetic fields, however, the electronic kinetic energy becomes
quenched by the creation of Landau levels \cite{kim} and Coulomb
interactions become important \cite{allan,Herbut06}. An alternative theory predicts that Coulomb interaction becomes relevant even at arbitrary small magnetic fields \cite{Herbut06}. Also, the minimal
conductivity\cite{Mish06} and collective excitations\cite{Hwang06}
depend on the electron-electron interaction.

In single graphene sheets, the unscreened electron-electron
interaction is - due to the vanishing density of state at the Dirac
point - marginally irrelevant in the renormalization group sense. This
leads to a behavior with the electronic self-energy of the form,
$\text{Im}\Sigma\propto\omega$, \cite{Gon96} which is reminiscent of
so-called marginal Fermi liquid behavior \cite{note}. The fact that
even the unscreened Coulomb interactions are marginally irrelevant in
single layer graphene suppresses the presence of magnetic phases in
this system. In fact, ferromagnetism can only be found in strong
coupling, when graphene's fine structure constant, $\alpha_G =
e^2/(\epsilon_0 \hbar v_F)$ ($e$ is electric charge, $\epsilon_0$ is
the dielectric constant, and $v_F$ is the Fermi-Dirac velocity), is
larger than a critical value $\alpha_c$. It was found that $\alpha_c
\approx 5.3$ for clean graphene ($\alpha_c \approx 3.8$ for disordered
graphene) \cite{Peres05}. For $\epsilon_0 = 1$ we have $\alpha_G
\approx 2$ which is smaller than the critical value indicating the
absence of a ferromagnetic transition in single layer graphene. For
Coulomb effects in disordered graphene, see also
Refs. \onlinecite{Sta05}.

In this paper, we study the electron-electron interaction in the
biased bilayer system (or unbiased graphene bilayer see
Ref. \cite{Wang07}). We use the standard tight-binding description of
the electronic structure \cite{McClure} and treat the problem in the
metallic regime when the chemical potential lies in the conduction
band leading to a topologically non-trivial ring for the Fermi
surface.  We also disregard inter-band transitions by assuming that
the electronic gap is sufficiently large so that the valence band is
completely filled and inert. We show that the ferromagnetic
instability can occur in this system due to the reduced phase space of
the ring.  We also calculate the polarization properties of this
system and show that they are rather unusual at high frequencies with
peculiar Friedel oscillations at low frequencies.

The paper is organized as followed. In section \ref{sec_model}, we introduce
the model and discuss the self-energy effects due to electron-electron
interactions. In section \ref{sec_ground_state}, we present a discussion
of the stability of the system towards
ferromagnetism by introducing a Landau-Ginzburg functional for the free
energy.  In section \ref{sec_susc}, we calculate two-particle correlation
functions, i.e., the imaginary part of the polarization and the plasmon
dispersion. We close with conclusions and future research directions.  

%%%%%%%%%%%%%%%%%%%%%%%%%%%%%%%%%%%%%%%%%%%%%%%%%%%%%%%%%%%%%%%%%%%
% Section : The model
%%%%%%%%%%%%%%%%%%%%%%%%%%%

\section{The effective model}
\label{sec_model}

\begin{figure}
\begin{center}\includegraphics*[%
    scale=0.4]{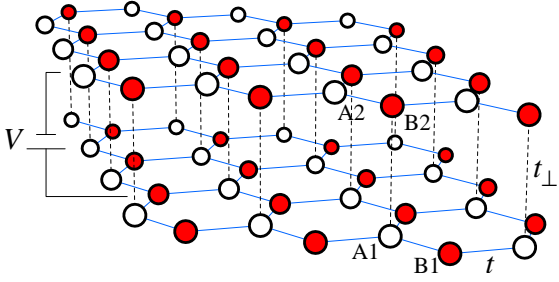}\end{center}
\caption{\label{cap:bilayer}(Color online) Lattice structure of a bilayer
graphene. The~A and~B sublattices are indicated 
by white and red spheres, respectively.}
\end{figure}

\begin{figure}
\begin{center}
\includegraphics[%
 scale=0.45]{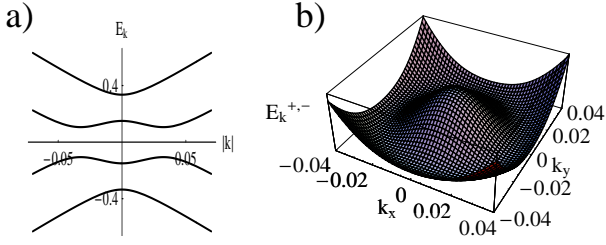}\end{center}
\caption{\label{fig_dispersion} a) The four bands $E_{\k}^{\pm,\pm}$ (in eV)
  as function of $|\k|$ (in \AA${}^{-1}$) around the $K$ point with
  $V=t_\perp$. b) The ``mexican-hat'' $E_{\k}^{+,-}$ (in eV) as function of
  $\k$ (in \AA${}^{-1}$) around the $K$ point.} 
\end{figure}

\begin{figure}
\begin{center}
\includegraphics[%
 scale=0.35]{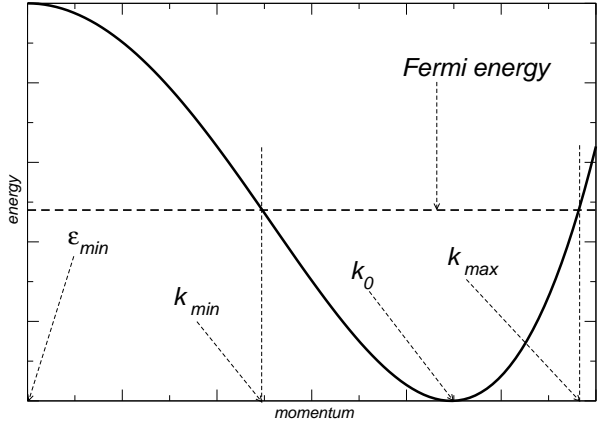}\end{center}
\caption{\label{fig_mexican} ``Mexican-hat'' dispersion of the biased
graphene bilayer close to the $K$ point in the Brillouin zone. The
symbols are explained in the text. }
\end{figure}

The graphene bilayer consists of two graphene planes (labeled $1$ and $2$), with a honeycomb
lattice structure, and stacked according to Bernal order (A1-B2), 
where~A and~B refer to each sublattice of a 
single graphene (see Fig.~\ref{cap:bilayer}).  
The system is parameterized by a tight-binding model
for the $\bm\pi$-electrons with nearest-neighbor 
in-plane hopping $t$ ($\approx 2.7$ eV)  and inter-plane hopping $t_\perp$
($\approx 0.3$ eV). The two layers have different
electrostatic potentials parameterized by $V$. 
The electronic band structure of the biased bilayer is obtained
within the tight-binding description \cite{eduardo}, leading to
the dispersion relation (we use units such that $\hbar = 1 = k_B$):
\begin{equation} \label{eq:Ekbias}
E_{\mathbf{k}}^{\pm\pm}(V) \!\!=\!\!
\pm \sqrt{\!\epsilon_{\mathbf{k}}^{2}  \!+\! t_{\perp}^{2}/2 \!+\!
V^{2}/4 \!\pm\! \sqrt{\!t_{\perp}^{4}/4
+(t_{\perp}^{2}\!+\!V^{2}\!)\epsilon_{\mathbf{k}}^{2}}}\,,
\end{equation}
where $\epsilon_{\mathbf{k}}$ is the dispersion of a single graphene layer,
\begin{eqnarray}
\epsilon_{\mathbf{k}} = t [3 + 2\cos(ak_{x}) + 
4\cos\left(ak_{x}/2\right)\cos\left(ak_{y}\sqrt{3}/2\right)]^{1/2} \, ,
\end{eqnarray}
where ${\bf k}=(k_x,k_y)$ is the two-dimensional (2D) momentum.
This dispersion relation around the $K$ and $K'$ points of the Brillouin
zone are shown in Fig. \ref{fig_dispersion}. Notice that the electronic
spectrum shows a ``mexican-hat'' dispersion with a gap minimum, 
$\Delta_g$, given by:  
\begin{eqnarray}
\Delta_g=
V[t_{\perp}^{2}/(t_{\perp}^{2} + V^2)]^{1/2} \, ,
\end{eqnarray}
that is located at momentum $k_0$ relative to the $K$ point. 
Close to the $K$ point the electronic dispersion can be written as:
\begin{eqnarray}
E(\bm k) \approx \Delta - \alpha k^2 + \lambda k^4 \, ,
\label{eq_mexican}
\end{eqnarray}
where
\begin{eqnarray}
\Delta &=& \frac{V}{2} \, ,
\nonumber
\\
\alpha &=& \frac {V}{t_\perp^2} v_F^2\, ,
\nonumber
\\
\lambda &=& 
V\left(
\frac {(t^2_\perp + V^2)^2}{V^2t^6_\perp}-\frac 1 {t^4_\perp}
\right)v_F^4 \, ,
\end{eqnarray}
where $v_F =  3 t a/2$ where $a=1.41$\AA${}$ is the lattice-spacing.
The energy dispersion (\ref{eq_mexican}) has the 
shape shown in Fig.~\ref{fig_mexican}, 
with its minimum at momentum
\begin{eqnarray}
k_0=\sqrt{\alpha/(2\lambda)} \, .
\end{eqnarray} 
For a finite density of electrons in the conduction band the occupied
momentum states are constrained such that $k_{\rm min,\sigma} < k < k_{\rm max,\sigma}$
where: 
\begin{eqnarray}
k^2_{\rm min, \sigma} &=& \frac{\alpha}{2\lambda}-2\pi n_\sigma\,,
\label{eq_kmin}
\\
k^2_{\rm max, \sigma} &=& \frac{\alpha}{2\lambda}+2\pi n_\sigma\,,
\label{eq_kmax}
\end{eqnarray}
where $n_\sigma$ is the electronic density per unit area
for electrons with spin $\sigma = \pm 1$. 

The density of states per unit area $\rho(\epsilon)$ can be written as:
\begin{equation}
\rho(\epsilon)=\frac 1 {4\pi\sqrt\lambda}\frac 1 {\sqrt {\epsilon-
\epsilon_{\rm min} } }\,,
\end{equation}
with $\epsilon_{\rm min}=\Delta -\alpha^2/(4\lambda)$, which has a square
root singularity at the band edge, just like in one-dimensional (1D) systems. 
This 1D singularity in the density of states of the biased bilayer leads to
an strong enhancement in the electron scattering either by other electrons
or impurities (which is beyond the scope of this paper). 
The Fermi energy, $\epsilon_F$,  can be expressed in terms of the electronic
density per unit area, $n_\sigma$, as
\begin{equation}
\epsilon_F=\epsilon_{\rm min} + (2\pi n_\sigma\sqrt\lambda)^2\,.
\end{equation}
%%%%%%%%%%%%%%%%%%%%%
%  Figure           % 
%%%%%%%%%%%%%%%%%%%%%
%%%%%%%%%%%%%%%%%%%%%

The simplest model including the effect of electron-electron interactions
in this problem is:
\begin{align}
H &= \sum_{\bm k,\sigma}E(\bm k)c^\dag_{\bm k,\sigma}c_{\bm k,\sigma}\\
&+
\frac 1 {2A}\sum_{\bm k, \bm p,\bm q}\sum_{\sigma,\sigma'}
V(q)
c^\dag_{\bm k+\bm q,\sigma}c^\dag_{\bm p-\bm q,\sigma'}
c_{\bm p,\sigma'}c_{\bm k,\sigma}\,,\notag
\end{align}
where $c_{\bm k,\sigma}$ ($c^{\dag}_{\bm k,\sigma}$) annihilates (creates)
and electron with momentum $\bm k$ and spin $\sigma$. Notice that 
only the Coulomb interaction between the electrons of the occupied
conduction band is considered.
The Fourier transform of the Coulomb potential is given by:
\begin{equation}
V(q)=\frac {2\pi e^2}{\epsilon_0 q}\,.
\end{equation} 
The unperturbed electronic Green's function reads:
\begin{equation}
G^0(\bm k,i\omega_n) = [i\omega_n - E(\bm k)]^{-1}\,,
\end{equation}
where $\omega_n=(2n+1)\pi/\beta$ are fermionic Matsubara frequencies.
%%%%%%%%%%%%%%%%%%%%%%%%%%%%%%%%%%%%%%%%%%%%%%%%%%%%%%%%%%%%%%%%%%%
% Subsection : First order Fermi liquid corrections
%%%%%%%%%%%%%%%%%%%%%%%%%%

\subsection{First order corrections to the quasi-particle spectrum}
%%%%%%%%%%%%%%%%%%%%%%%%%%%%%%%%%%%%
%%%%%%%%% Figure
%%%%%%%%%%%%%%%%%%%%%%%%%%%%%%%%%%%%
\begin{figure}
\begin{center}
\includegraphics[%
 scale=0.35]{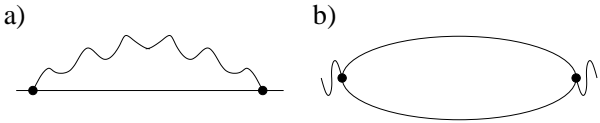}\end{center}
\caption{\label{fig_diagrams} Evaluated diagrams where the solid line
  resembles the bare electronic propagator and the wiggled line the bare
  Coulomb interaction: a) electronic self-energy. b) particle-hole bubble or
  polarization.} 
\end{figure}
%%%%%%%%%%%%%%%%%%%%%%%%%%%%%%%%%%%%
The first order correction to the electronic self-energy
is given by the diagram in Fig. \ref{fig_diagrams} (a),
and it reads:
\begin{equation}
\Sigma_1(\bm k)=-\frac {2\pi e^2}{\epsilon_0 A}\sum_{\bm q}
\frac 1{\beta}
\sum_{n}
\frac 1 {q}G^0(\bm k-\bm q,i\omega_n)\,,
\end{equation}
and considering the zero temperature limit, one gets
\begin{eqnarray}
\Sigma_1(\bm k)&=&
-\frac { e^2k_{\rm max}}{2\pi\epsilon_0}I_1(a,k/k_{\rm max})\,,
\\\notag
&\simeq& -\frac { e^2k_{\rm max}}{2\pi\epsilon_0}\Big[
2(1-a)\pi +\frac 1 2 \left(\frac 1 a -1\right)\pi \frac {k^2}
{k^2_{\rm max}}\\\notag
&&\quad\quad\quad+\frac 3 {32} \left(\frac 1 {a^3}-1 \right)\pi
\frac {k^4}
{k^4_{\rm max}}
\Big]\;,
\end{eqnarray}
where $a=k_{\rm min}/k_{\rm max}$ and the Coulomb integral $I_1(x)$ is given
in appendix \ref{app_Coulomb}. From this result one obtains the renormalized values for $\Delta$, $\alpha$ and $\lambda$,
defined as $E_1(\bm k)=E(\bm k)+\Sigma_1(\bm k)$.
For $k_{\rm min}>0$, i.e., the Mexican hat is partially filled, the renormalized values read
\begin{align}
\Delta_1 &= \Delta  
-\frac { e^2}{\epsilon_0}
(k_{\rm max}-k_{\rm min})\,,
\label{eq_d1}\\
\alpha_1 &= \alpha +\frac { e^2}{4\epsilon_0}
\left(\frac {k_{\rm max}-k_{\rm min}}{k_{\rm max}k_{\rm min}} \right)\,,
\label{eq_a1}\\
\lambda_1 &= \lambda -\frac { 3e^2}{64\epsilon_0}
\left(\frac {k^3_{\rm max}-k^3_{\rm min}}{k^3_{\rm max}k^3_{\rm min}} \right)
\,.
\label{eq_l1}
\end{align}
A consequence of the renormalization 
of the bare parameters is that for densities such that $\lambda_1\leq0$, the
spectrum becomes unbounded indicating the a possible instability of the
non-interacting system. For realistic values of the parameters (see Section
\ref{sec_ground_state}), the instability occurs at $k_{\rm min}/k_0\approx3/4$, i.e.,
for densities close to $n_\sigma\approx n_\sigma^c=k_0^2/(2\pi)$ when the
Mexican hat is completely filled.  
We will confine ourselves to the limit of low densities so that the expansion
of the energy dispersion in Eq. (\ref{eq_mexican}) up to
the quartic term in the electron dispersion is valid. For larger densities
for which $\lambda_1\approx0$, higher order terms in the expansion must
be included. For $n_\sigma\geq n_\sigma^c$, i.e., the Mexican hat is completely filled,
the electron-electron interaction stabilizes the dispersion and the dip in
the spectrum reduces. In this case, the parameters of the model modify to:
\begin{align}
\Delta_1 &= \Delta-\frac { e^2}{\epsilon_0}k_{\rm max}\;,\\
\alpha_1 &= \alpha -\frac { e^2}{4\epsilon_0}k_{\rm max}^{-1}\;,\\
\lambda_1&=\lambda+\frac { 3e^2}{64\epsilon_0}k_{\rm max}^{-3}\;.
\end{align}

%%%%%%%%%%%%%%%%%%%%%%%%%%%%%%%%%%%%%%%%%%%%%%%%%%%%%%%%%%%%%%%%%%%
% Section : on the nature of the ground state
%%%%%%%%%%%%%%%%%%%%%%%%%%%%%%%%%%%%%%%%%%%%%%%%%%%%%%%%%%%%%%%%%%%
\section{Exchange instability of the ground state}
\label{sec_ground_state}

In this section we examine a possible instability of the Fermi ring is
unstable towards a ferromagnetic ground state.
For the instability to occur, the exchange energy due to Coulomb interactions
has to overcome the loss of kinetic energy when the system is polarized.
The kinetic energy for electrons of spin $\sigma$ has the form:
\begin{equation}
K_\sigma (n,m)=\frac {A}{2\pi}\left[\left(
 2\pi\Delta - \frac {\pi}{2}\frac{\alpha^2}{\lambda}
\right)n_\sigma + 
\frac 8 3 \lambda\pi^3n_\sigma^3
\right]\;,
\label{kinetic}
\end{equation}
where $A$ is the area of the bilayer. 
The exchange energy for electrons with spin $\sigma$ can be written as
\begin{align}
E_\sigma^{\rm Exc}=\frac{1}{2A}\sum_{\rm k,\rm k'\text{ occup.}}V(|\rm k-\rm k'|)|\langle\rm k|\rm k'\rangle|^2\;.
\end{align}
 Within the approximations described in the introduction, i.e., assuming that the wave functions are mostly localized on one sublattice in one layer, the overlapp factor $\langle\rm k|\rm k'\rangle\approx1$. This leads to 
\begin{equation}
E_\sigma^{\rm Exc}(n,m)=-\frac {e^2A}{8\pi^2\epsilon_0}
k^3_{{\rm max},\sigma}\,I_2
\left(\frac {k_{{\rm min},\sigma}}{k_{{\rm max},\sigma}}
\right) \,,
\label{exchange}
\end{equation}
where $k_{{\rm max},\sigma}$ and $k_{{\rm min},\sigma}$
are defined in Eqs. (\ref{eq_kmin}) and (\ref{eq_kmax}),
(see Fig. \ref{fig_mexican}). 

In the paramagnetic phase both spin bands have the same
number of electrons and hence $n_\sigma=n/2$. In the ferromagnetic phase
the system acquires a magnetization density, $m=n_+ - n_-$, and the electron occupation
change to: $n_\sigma = (n+\sigma m)/2$.  
In order to study a possible ferromagnetic transition, we 
derive a Landau-Ginzburg functional in powers of $m$,
with the coefficients depending on $n$.
The Landau-Ginzburg functional is defined as:
\begin{align}
F[\alpha,\lambda,n,m]&= \sum_{\sigma}[
K_\sigma (n,m)+
E_\sigma^{\rm Exc}(n,m)
\\\nonumber
&\quad\quad-K_\sigma (n,0)-
E_\sigma^{\rm Exc}(n,0)]\nonumber\\\nonumber
&\simeq a(n)m^2+b(n)m^4+c(n)m^6+\ldots\,. 
\end{align} 
A second order magnetic transition occurs when $a(n)$ becomes negative.
If $a(n)$ is always positive, a first order transition
may occur if $b(n)$ becomes negative. 
The  term linear in $n_\sigma$ in the kinetic energy (\ref{kinetic})
does not contribute to $F[\alpha,\lambda,n,m]$.

The coefficient $a(n)$ has two main contributions:
one comes from the kinetic energy (\ref{kinetic}), $a^K(n)$, and another arises from
the exchange (\ref{exchange}), $a^{\rm ex}(n)$. These contributions read:
\begin{align}
a^K(n) &= A\lambda \pi^2n\,,\\
a^{\rm ex}(n) &= -\frac {e^2A}{4\epsilon_0}(L_0+L_1+L_2)\,,
\end{align}
where the coefficients $L_i$ are functions related to the Coulomb integral
$I_2(x)$ defined in appendix \ref{app_Coulomb} in the following way: 
\begin{align}
\notag
L_0&= \frac 3 8 k^{-1}_{\rm max}I_2
\left(\frac {k_{\rm min}}{k_{\rm max}}
\right)\\\notag
L_1&=-\frac 3 2 \frac {k^2_0}{k^2_{\rm max}\,k_{\rm min}}
I^{'}_2
\left(\frac {k_{\rm min}}{k_{\rm max}}
\right)\\\notag
L_2&=\frac {k_0^4}{2k^3_{\rm max}\,k^2_{\rm min}}
I^{''}_2
\left(\frac {k_{\rm min}}{k_{\rm max}}
\right)
+
\frac {k_0^2(k^2_0-2n\pi)}{2k^2_{\rm max}\,k^3_{\rm min}}
I^{'}_2
\left(\frac {k_{\rm min}}{k_{\rm max}}
\right)
\end{align}

%%%%%%%%%%%%%%%%%%%%%
%  Figure           % 
%%%%%%%%%%%%%%%%%%%%%
\begin{figure}[t]
\begin{center}
\includegraphics[angle=0,width=1.\linewidth]{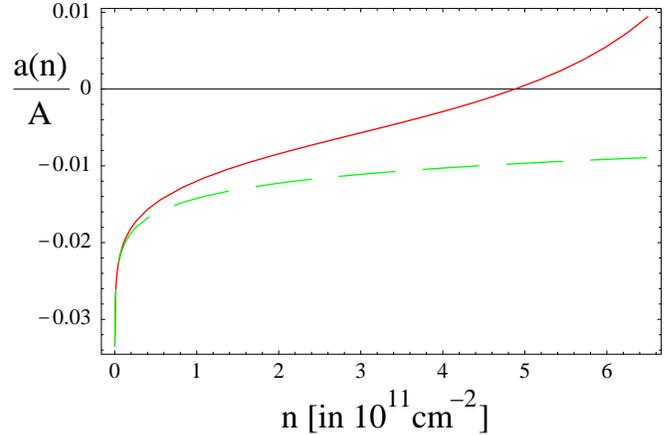}\end{center}
\caption{\label{fig_CA} (Color online): Plot of $a(n)/A$ (in units of
eV$[10^{11}\text{cm}^{-2}]^{-1}$), as function of the electronic density
$n$ (in units of $10^{11}\text{cm}^{-2}$).  The dashed (green) line gives the
bare value of $a(n)/A$, the full (red) line contains self-energy
corrections. }
\end{figure}

%%%%%%%%%%%%%%%%%%%%%
%  Figure           % 
%%%%%%%%%%%%%%%%%%%%%
\begin{figure}[t]
\begin{center}
\includegraphics[angle=0,width=1.\linewidth]{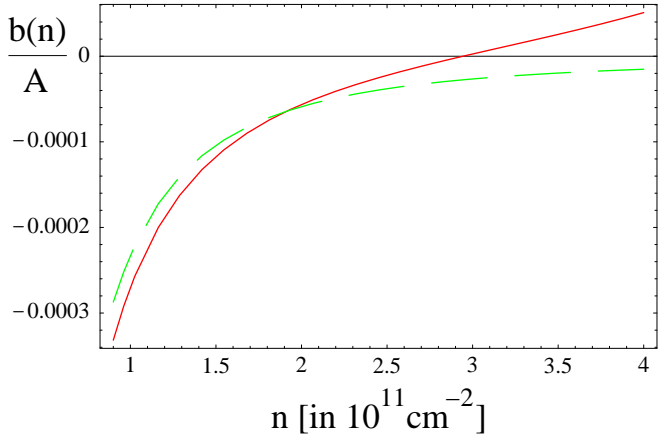}\end{center}
\caption{\label{fig_CB} (Color online): Plot of $b(n)/A$ (in units of
eV$[10^{11}\text{cm}^{-2}]^{-3}$), as function of the electronic density
$n$ (in units of $10^{11}\text{cm}^{-2}$).  The dashed (green) line
gives the bare value, the full (red) line is the result with
self-energy corrections.}
\end{figure}

%%%%%%%%%%%%%%%%%%%%%
%  Figure           % 
%%%%%%%%%%%%%%%%%%%%%
\begin{figure}[t]
\begin{center}
\includegraphics[angle=0,width=1.\linewidth]{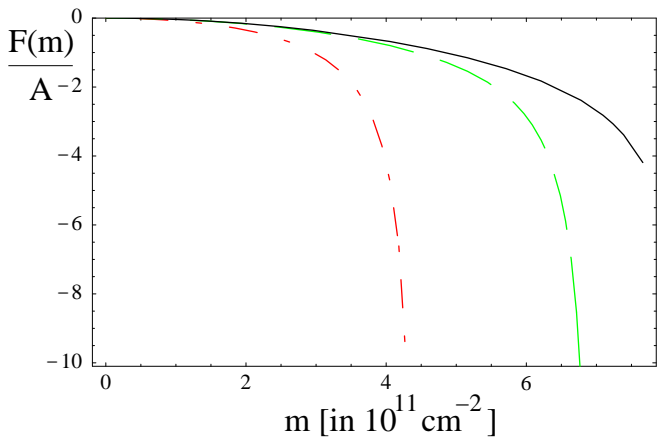}\end{center}
\caption{\label{fig_Fren} (Color online): Plot of $F(m)/A$ (in units
of eV$[10^{11}\text{cm}^{-2}]$), as function of the magnetic density
$m$ (in units of $10^{11}\text{cm}^{-2}$) for various electronic
densities $n$ including self-energy corrections. The dotted-dashed
(red) line stands for $n=5\times10^{11}\text{cm}^{-2}$, the dashed (green)
line for $n=2.5\times10^{11}\text{cm}^{-2}$ and the full (black) line for
$n=10^{11}\text{cm}^{-2}$.}
\end{figure}
%%%%%%%%%%%%%%%%%%%%%

In the following, we choose the parameters as in section
\ref{sec_model} with $V=t_\perp$. For $\alpha_G\approx2$ all coefficients
are negative indicating that the system has a strong tendency towards
ferromagnetic ordering.  In this case the transition is of second
order.  There are, however, values of $\alpha_G$ for which $a(n)$ is
positive, whereas the remaining coefficients are negative. In this
case the transition is of first order. At zero temperature the
magnetization density equals the density of electrons.

We have also investigated how the above picture changes if one
includes the corrections to energy spectrum due to the
electron-electron interaction, given by Eqs. (\ref{eq_a1}) to
(\ref{eq_l1}). The comparison is shown in Fig. \ref{fig_CA} and Fig.
\ref{fig_CB} where the dashed line shows the coefficients without
self-energy effects whereas the full line includes self-energy
effects. The inclusion of self-energy effects decreases the tendency
toward a ferromagnetic ground-state and there is a critical density
for which both coefficients, $a(n)$ and $b(n)$, become positive.  This
is also seen in Fig. \ref{fig_Fren}, where the free energy $F(m)$ is
shown for various electronic densities. Including self-energy effects
leads to the saturation of the magnetization at high densities, i.e.,
for $n=5\times10^{11}\text{cm}^{-2}$ the magnetization is not given by
$m=n$ but by a lower value. Nevertheless, at low densities
($n\leq10^{11}\text{cm}^{-2}$) the picture is not modified by
self-energy corrections and we expect ferromagnetic ordering at $T=0$.

%%%%%%%%%%%%%%%%%%%%%%%%%%%%%%%%%%%%%%%%%%%%%%%%%%%%%%%%%%%%%%%%%%%
% Sub Section : Second order corrections
%%%%%%%%%%%%%%%%%%%%%%%%%%%

\section{Polarization}
\label{sec_susc}

In this Section, we calculate the bilayer density-density correlation function,
i.e., the response of the system to an external potential. The bare
polarization as a function of momentum $q$ and frequency $\omega$ 
is given in terms of the bare particle-hole bubble shown in
Fig. \ref{fig_diagrams} (b). It reads:  
\begin{align}
\notag
P^{(1)}(\q,\omega)&=\frac{2}{(2\pi)^2}\int d^2k\frac{n_F(E(\bm k))-n_F(E(\bm
  k+\bm q))}{E(\bm k)-E(\bm k+\bm q)-\omega - i\delta},
\end{align}
where $n_F(E)$ is the Fermi-Dirac function.
Hence, the imaginary part of the retarded response function is given by:
\begin{align}
\text{Im}P^{(1),\text{ret}}(\q,\omega)&=\frac{1}{2\pi}\int d^2kn_F(E(\bm k))\\\notag
&\times\sum_{\xi=\pm}\xi\delta(\xi\omega-(E(\bm k)-E(\bm k+\bm q)))\;.
\end{align}

Assuming a small electronic density and hence electrons with momentum $k$ 
such that $|k-k_0| \ll k_0$, the energy dispersion can be approximated as:
\begin{align}
E(\bm k)\approx(|\bm k|-1)^2\;,
\label{dispersion}
\end{align}
where we have set $k_0=1$ (thus we measure momentum in units of $k_0$) and
take $\alpha k_0^2$ as our unit of energy (only in this
section). The polarization is an odd function of $\omega$ and hence we can 
consider only the case of $\omega>0$.  
Defining the function:
\begin{align}
f_{\xi}(x)&\equiv\xi\omega-(E(\bm k)-E(\bm k+\bm q))\notag\\
&=\xi\omega+q^2+2k+2kqx-2\sqrt{k^2+q^2+2kqx}\notag \, ,
\end{align}
with $\xi=\pm1$ and its zeros:
\begin{align}
\notag
x_{\xi}^\pm=\frac{1}{2kq}\left[\left(\sqrt{-\xi\omega+(k-1)^2}\pm1\right)^2-k^2-q^2\right]
\, ,
\end{align}
the imaginary part of the polarization can be written as:
\begin{align}
\label{Polarization_kINT}
\text{Im}P^{(1),\text{ret}}(\q,\omega)&=\frac{2}{2\pi}
\int_{1-b}^{1+b} dkk\sum_{\xi,\gamma=\pm}\xi\\\notag
&\times\frac{1}{|f_{\xi}'(x_{\xi}^\gamma)|}\frac{1}
{\sqrt{1-(x_{\xi}^\gamma)^2}}\Theta(1-(x_{\xi}^\gamma)^2)\;,\notag 
\end{align}
where $b\ll1$ defines the Fermi sea, i.e., $b=(k_{\rm max}-k_{\rm
min})/2$ and no explicit spin-dependency is considered. The integral
in (\ref{Polarization_kINT}) can be evaluated analytically but the
expressions are lengthy. Here we focus on the limits of high- and
low-frequency relative to the Fermi-energy, $\epsilon_F\sim b^2$. In
the low-frequency limit, we further consider the cases of $|\bm
q|\ll1$ and $|\bm q-2|\ll1$, representing forward and backward
scattering processes with small momentum transfer, respectively.

For the analytical approximations, it is convenient to perform separate
substitutions for $\xi=+1$ and $\xi=-1$. For $\xi=-1$, we set:
\begin{align}
(k-1)=\sqrt{\omega}\sinh\varphi\;,
\end{align}
where the upper and lower bounds in the integral are given by $+\varphi_0$
and $-\varphi_0$ , respectively,
where $\varphi_0=\sinh^{-1} b/\sqrt{\omega}$. We thus obtain 
\begin{widetext}
\begin{align}
\text{Im}P^{(1),\text{ret},\xi=-1}(\q,\omega)&=-\frac{2}{2\pi}
\int_{D}d\varphi\sum_{\gamma=\pm}\sqrt{\omega}
\cosh\varphi(\sqrt{\omega}\sinh\varphi+1)
\frac{|1+\gamma\sqrt{\omega}\cosh\varphi|}{|1-|1+\gamma\sqrt{\omega}\cosh\varphi||}\frac{1}{\sqrt{D_1D_2}}\;, 
\end{align} 
\end{widetext}
where we abbreviate the denominators as,
\begin{eqnarray}
\notag
D_1&=-&(\sqrt{\omega}\sinh\varphi+1-q)^2+(1+\gamma\sqrt{\omega}\cosh\varphi)^2\;,\\
\notag
D_2&=&(\sqrt{\omega}\sinh\varphi+1+q)^2-(1+\gamma\sqrt{\omega}\cosh\varphi)^2\;.
\end{eqnarray}
The final result must be real in such a way that there might be a constraint
in the integration domain, denoted by $D$.

For $\xi=1$, we only have a contribution if $\omega<(k-1)^2\leq b^2$ and we can perform the substitution, 
\begin{align}
|k-1|=\sqrt{\omega}\cosh\varphi\;.
\end{align}
The upper integration limit is given by $\varphi_0=\cosh^{-1}b/\sw$.
We thus obtain:
\begin{widetext}
\begin{align}
\text{Im}P^{(1),\text{ret},\xi=1}(\q,\omega)&=
\frac{2}{2\pi}\int_{D;\varphi\geq0}d\varphi
\sum_{\gamma=\pm}\sqrt{\omega}\sinh\varphi\sum_{\zeta=\pm}(\zeta\sqrt{\omega}\cosh\varphi+1)
\frac{|1+\gamma\sqrt{\omega}\sinh\varphi|}{|1-|1+
\gamma\sqrt{\omega}\sinh\varphi||}\frac{1}{\sqrt{D_1D_2}}\;, 
\end{align}
\end{widetext}
where we abbreviate the denominators as,
\begin{eqnarray}
\notag
D_1&=-&(\zeta\sqrt{\omega}\cosh\varphi+1-q)^2+(1+\gamma\sqrt{\omega}\sinh\varphi)^2\;,\\
\notag
D_2&=&(\zeta\sqrt{\omega}\cosh\varphi+1+q)^2-(1+\gamma\sqrt{\omega}\sinh\varphi)^2\;.
\end{eqnarray}

\subsection{High frequency}

\begin{figure}[t]
\begin{center}
\includegraphics[angle=0,width=1.\linewidth]
{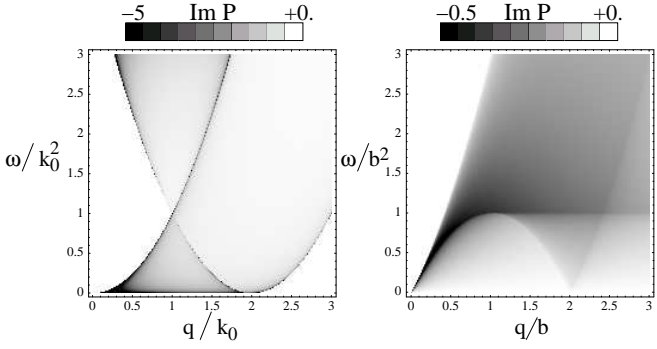}\end{center}
\caption{\label{fig_Forward} 
Left hand side: Density plot of the imaginary part of the polarization,
$\text{Im}P^{(1),\text{ret}}(\q,\omega)/b$, as function of $q$ and $\omega$
in units of $k_0$ for $b\to0$. Right hand side: 
Density plot of the imaginary part of the susceptibility,
$\text{Im}P^{(1),\text{ret}}(\q,\omega)$, as function of $q$ and $\omega$ in
units of $b\ll k_0$.} 
\end{figure}

At high frequencies, $\omega\gg b^2$, we can expand $\sinh\varphi\sim\varphi$,
$\cosh\varphi\sim1$ and the resulting integral can be performed.  
Both denominators, $D_1$ and $D_2$, have to be real, altering the
integration domain. But for $b\to0$, the lower and upper bounds are
given by $\pm b/\sw$ or zero (since the crossover region is
missing). In this approximation, the integrand can thus be expanded in
$\varphi$ and only the constant term can be kept. We obtain: 
\begin{widetext}
\begin{align}
\text{Im}P^{(1),\text{ret}}(\q,\omega)&=-(2b)\frac{2}{2\pi}
\sum_{\gamma=\pm}\frac{|1+\gamma\sqrt{\omega}|}{|1-|1+\gamma\sqrt{\omega}||}
\text{Re}\frac{1}{\sqrt{(1+q)^2-(1+\gamma\sw)^2}}\text{Re}
\frac{1}{\sqrt{-(1-q)^2+(1+\gamma\sw)^2}} 
\,  . 
\end{align}
\end{widetext}
The restrictions, i.e., the regions in which $\text{Im}P^{(1),\text{ret}}=0$,
are obtained by imposing real denominators. The results are shown on the left
hand side of Fig. \ref{fig_Forward}.  

For $\omega\to0$, we obtain:
\begin{align}
\label{HighEnergyNFL}
\frac{\text{Im}P^{(1),\text{ret}}(\q,\omega)}{-(\pi/2bk_0)} =
\left\{\begin{array}{l l}
\omega^{-1/2}/(q\sqrt{1-(q/2k_0)^2})&;\;0<q<2b\\
0&;\;q\geq2b
\end{array}\right.
\end{align}
We thus observe a pronounced non-Fermi liquid behavior in the high-energy
regime, $b^2 \le \omega \le k_0^2$. Only in the small energy regime
$\omega<b^2$, Fermi-liquid behavior 
is recovered (see next section).
  
\subsection{Low frequency; forward scattering}

In the low-frequency regime, $\omega\sim b^2$, we limit ourselves to forward
scattering processes with $q\ll1$.  
Only considering the lowest order of $\sqrt\omega\sim q$, we obtain for
$\xi=-1$ the following approximate denominators: 
\begin{align}
\notag
D_1\to2q+\gamma2\sw e^{-\gamma\varphi}\;,\;D_2\to2q-\gamma2\sw
e^{-\gamma\varphi} \, ,
\end{align}
yielding, 
\begin{align}
\text{Im}&P^{(1),\text{ret},\xi=-1}(\q,\omega)\\\notag
&\quad\quad=-\frac{1}{2\pi}\int_{-\sinh^{-1}\tb}^{\text{min}(\ln
  \tq,\sinh^{-1}\tb)}d\varphi\frac{1}{\sqrt{\tq^2-e^{2\varphi}}}\quad, 
\end{align}
where we defined $\tb=b/\sw$ and $\tq=q/\sw$.

For $\xi=1$, and again only considering the lowest order of $\sqrt\omega\sim
q$, we obtain the following approximate denominators: 
\begin{align}
\notag
D_1\to2q-\zeta2\sw e^{-\gamma\zeta\varphi}\;,\;
D_2\to2q+\zeta2\sw e^{-\gamma\zeta\varphi} \, ,
\end{align}
leading to, 
\begin{align}
\text{Im}&P^{(1),\text{ret},\xi=1}(\q,\omega)\\\notag
&\quad\quad=\frac{1}{2\pi}\int_{0}^{\text{min}(\ln
  \tq,\cosh^{-1}\tb)}d\varphi\frac{1}{\sqrt{\tq^2-e^{2\varphi}}}\\\notag 
&\quad\quad+\frac{1}{2\pi}\int_{\text{max}(-\ln
  \tq,0)}^{\cosh^{-1}\tb}d\varphi\frac{1}{\sqrt{\tq^2-e^{-2\varphi}}}\quad. 
\end{align}

The analytical expressions are presented in appendix \ref{app_susc}. The
behavior for $\omega\to0$ is given by ($P=-4\pi
q\text{Im}P^{(1),\text{ret}}(\q,\omega)$): 
\begin{align}
P=
\left\{\begin{array}{l l}
(\omega/b^2)&;\;0<q<2b\\
(\omega/b^2)^{1/2}/\sqrt{2}&;q=2b\\
(\omega/b^2)(1+1/\sqrt{1-(2b/q)^2})&;\;q>2b
\end{array}\right.
\end{align}
The results are shown on the right hand side of
Fig. \ref{fig_Forward}. One sees a linear mode for small wave vectors
$q\ll1$, reminiscent of 1D Luttinger liquids. Nevertheless, the
asymptotic behavior of the polarization for small energies shows
Fermi-liquid behavior for $q\neq 2b$. At $q=2b$, we find non-Fermi
liquid behavior since two parallel regions of the Fermi surface are
connected. The non-analyticity at $q=2b$ leads to Friedel oscillations
with period $\pi/b$ decaying as $r^{-2}$ at large distances. These
results are consistent with alternative analytical treatments for
generic Fermi surfaces with arbitrary curvature\cite{GGV97,FG02}.

\subsection{Low frequency; backward scattering}

For $\q\approx2\kNull$, i.e., for backward scattering processes, we consider
scattering processes with momentum $2+q$ and $|q|\ll1$. We then get for
$\xi=-1$ the following approximate denominators for low energies $\sw\sim q$: 
\begin{align}
D_1\to\mp2q+\gamma2\sw e^{\gamma\varphi}\quad,\quad D_2\to2 \, .
\end{align}
For $\xi=1$, we obtain,
\begin{align}
D_1\to\mp2q+\zeta2\sw e^{\gamma\zeta\varphi}\quad,\quad D_2\to2\;.
\end{align} 

The analytical solution is presented in appendix \ref{app_susc}. Defining
$P^1+P^2=-4\pi\text{Im}P^{(1),\text{ret}}(2\kNull^\pm,\omega)$ with
$|2\kNull^\pm|=2+q$, where $q$ can be positive or negative, we obtain the
following behavior for $\omega\to0$: 
\begin{align}
P^1=(\omega/b^2)\left[\frac{1}{2\sqrt{|q|}}+\frac{1}{2\sqrt{2+|q|}}\right]\Theta(-q)
\, ,
\end{align}
\begin{align}
P^2=
\left\{\begin{array}{l l}
(\omega/b^2)/2\sqrt{|q|-2b}&;\;q<-2b\\
(\omega/b^2)^{1/2}/\sqrt{2b}&;\;q=-2b\\
(\omega/b^2)/2\sqrt{|q|}&;\;-2b<q<0\\
(\omega/b^2)^{1/2}/\sqrt{2b}&;\;q=0\\
(\omega/b^2)/2\sqrt{2b-q}&;\;0<q<2b\\
(\omega/b^2)^{1/2}/\sqrt{2b}&;\;q=2b\\
0&;\;q>2b
\end{array}\right.
\end{align}

\begin{figure}[t]
\begin{center}
\includegraphics[angle=0,width=1.\linewidth]{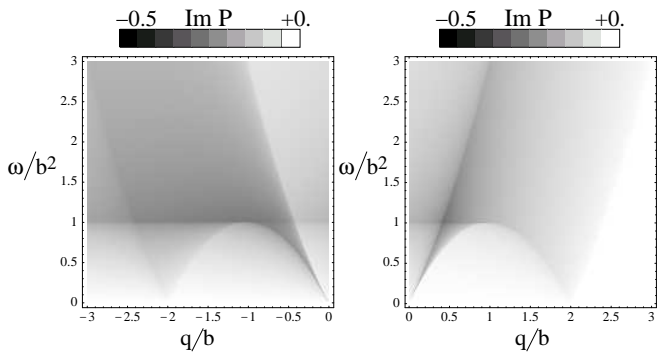}\end{center}
\caption{\label{fig_Backward} 
Density plot of the imaginary part of the polarization,
$\text{Im}P^{(1),\text{ret}}(2\kNull^\pm,\omega)$ with $|2\kNull^\pm|=2+q$,
as function of $q$ and $\omega$ in units of $b\ll k_0$. 
}
\end{figure}

The result for ``positive'' and ``negative'' back-scattering is shown on the
right and left hand side of Fig. \ref{fig_Backward}, respectively. 

There are three wave numbers that connect two parallel regions of the
Fermi surface and thus lead to non-Fermi liquid behavior. This results
in Friedel oscillations with period $\pi/k_0$ modulated by
oscillations with period $\pi/b$, both decaying as $r^{-2}$ at large
distances. These are enhanced by a factor $(2b)^{-1/2}$ compared to
the Friedel oscillations originating from forward scattering processes
with $q=2b$.

\begin{figure}[t]
\begin{center}
\includegraphics[angle=0,width=0.8\linewidth]{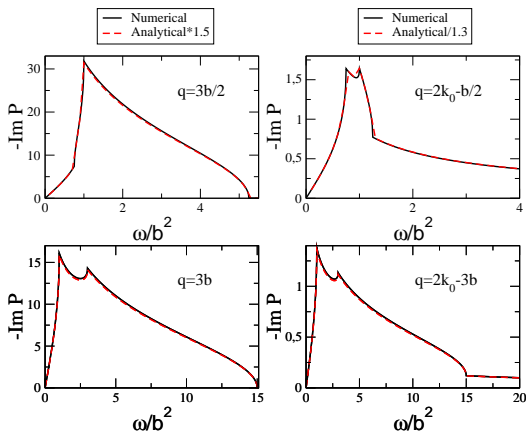}\end{center}
\caption{\label{fig_NFL} 
(Color online) The imaginary part of the polarization,
$\text{Im}P^{(1),\text{ret}}(q,\omega)$, for $q=3b/2, 3b$ (left hand side)
and $q=2k_0-b/2, 2k_0-3b$ (right hand side), as function of $\omega$. The
solid (black) line shows the numerical and the dashed (red) line the
analytical result multiplied by a global factor.} 
\end{figure}

We have integrated Eq. (\ref{Polarization_kINT}) numerically and obtain good
agreement between the numerical and the analytical result of appendix
\ref{app_susc} in the whole energy regime and for all wave numbers up to a
global factor $0.75$ and $1.3$ in case of forward and backward scattering
processes, respectively. In Fig. \ref{fig_NFL}, the numerical (full line) and
analytical (dashed line) results for $\text{Im}P^{(1),\text{ret}}$ are shown
for $q=3b/2, 3b$ (left hand side) and $q=2k_0-b/2,2k_0-3b$ (right hand side)
as function of with $\omega$. In all cases, the low-energy regime is
characterized by Fermi-liquid behavior whereas the high-energy shows the
square-root divergence predicted by Eq. (\ref{HighEnergyNFL}).    
%%%%%%%e%%%%%%%%%%%%%%%%%%%%%%%%%%%%%%%%%%%%%%%%%%%%%%%%%%%%%%%%%%%%
% SubSection : The plasmon spectrum
%%%%%%%%%%%%%%%%%%%%%%%%%%%

\subsection{The plasmon spectrum}

Within the random phase approximation (RPA), the electronic dielectric constant is given by:
\begin{align}
\epsilon(\q,\omega)=1-v_\q P^{(1),\text{ret}}(\q,\omega) \, ,
\end{align}
where $v_\q=2\pi e^2/|\q|$. The plasmon dispersion, $\Omega_q$, is then given
 by the zeros of the dielectric function, $\epsilon(\q,\Omega_q)=0$. 
In the long wavelength limit, $\q\to0$, we have:  
\begin{align}
P^{(1),\text{ret}}(\q\to0,\omega)=\frac{2q^2\alpha
  n}{\omega^2}=\frac{2q^2Vv_F^2n}{\omega^2t_\perp^2} \, .
\end{align}

Let us, at this point, recall the result for a 2D electron gas \cite{Stern}
where the dispersion is given by $E(\bm k)=k^2/(2m^*)$, where $m^*$ is the
effective mass. The polarization function is given by:
\begin{align}
P^{(1),\text{ret}}_{{\rm 2DEG}}(\q,\omega)&=\frac{nq^2}{m^*\omega^2}\quad , 
\end{align}
leading to the well-known result:
\begin{align}
\Omega_q=\left(\frac{2\pi ne^2}{m^*}q\right)^{1/2}\quad.
\end{align}

Coming back to the Fermi ring and introducing the effective mass
$m^*=(4\alpha)^{-1}$, we obtain the same expression if we include the
degeneracy factor $g_V=2$ for the two nonequivalent Fermi points in one
graphene layer. With the electrostatic relation $V=e^2nd$ ($d$ being the
distance between the two layers), we thus obtain 
\begin{align}
\Omega_q=\frac{ne^2v_F}{t_\perp}\sqrt{dq}\quad.
\end{align}

%%%%%%%%%%%%%%%%%%%%%%%%%%%%%%%%%%%%%%%%%%%%%%%%%%%%%%%%%%%%%%%%%%%
% Section : conclusions
%%%%%%%%%%%%%%%%%%%%%%%%%%%%%%%%%%%%%%%%%%%%%%%%%%%%%%%%%%%%%%%%%%%

\section{Conclusions}
\label{sec_conclusions}

We have analyzed the effect of electron-electron interactions in a biased
bilayer graphene in the regime where the Fermi surface has the shape of a
ring. We have studied the stability of this system towards ferromagnetic
order using a one-band model of the ``Mexican hat'' dispersion. We find 
that the spin polarized phase is stable. Unfortunately, our single band 
calculation does not allow the calculation of the
saturation magnetization, which we plan to study in the future, taking into
account the full band structure.

The unusual electronic occupation in $k$-space also leads to a deviations
from the predictions of Landau's theory of a Fermi liquid,
$\text{Im}P\sim|\omega|$ at low energies,
$(Vv_F^2/t_\perp^2)b^2 \ll \epsilon \ll(Vv_F^2/t_\perp^2)k_0^2$, (see
Fig.[\ref{fig_NFL}]).  The $\text{Im}P\sim|\epsilon|^{-1/2}$ dependence found
in this range translates into a quasiparticle lifetime which decays as
$\Gamma ( \epsilon ) \propto \sqrt{| 
\epsilon - \epsilon_F|}$\cite{RLTG06}. 

At low energies, $\epsilon \ll (Vv_F^2/t_\perp^2)b^2$, the system
resembles a Fermi liquid except for wave numbers which connect two
parallel regions of the Fermi surface, i.e., for $q=2b,2k_0$ and
$q=2k_0\pm2b$. Note that, as the Fermi velocity is proportional to the
width of the Fermi ring, $b$, the imaginary part of the response
function grows as $b^{-2}$.  The existence of two Fermi lines implies
Friedel oscillations with period $\pi/b$. The plasmon dispersion shows
typical features of two-dimensional systems, nevertheless the energy
scale is greatly enhanced compared to a 2DEG.

%%%%%%%%%%%%%%%%%%%%%%%%%%%%%%%%%%%%%%%%%%%%%%%%%%%%%%%%%%%%%%%%%%%
% Section : Acknowledgments
%%%%%%%%%%%%%%%%%%%%%%%%%%%

\subsection*{Acknowledgments}
This work has been supported by MEC (Spain) through Grant
 No. FIS2004-06490-C03-00, by the European Union, through contract
 12881 (NEST), and the Juan de la Cierva Program (MEC,
 Spain). N.~M.~R.~Peres thanks the ESF Science Programme INSTANS
 2005-2010, and FCT under the grant PTDC/FIS/64404/2006. A.~H.~C.~N
 was supported through NSF grant DMR-0343790.

%%%%%%%%%%%%%%%%%%%%%%%%%%%%%%%%%%%%%%%%%%%%%%%%%%%%%%%%%%%%%%%%%%%
% Section : Appendix
%%%%%%%%%%%%%%%%%%%%%%%%%%%
\appendix
\section{Coulomb integrals}
\label{app_Coulomb}
In this appendix, we list the results for the two relevant Coulomb
integrals. The self-energy corrections involve the following
Coulomb-integral: 
\begin{align}
I_1(a,y)&=\int_0^{2\pi}\int_a^1\frac {x dx d\theta}
{\sqrt{x^2+y^2\pm 2xy\cos\theta}}\\\notag&=
2(1+y)\bm E[g(y)] + 2(1-y)\bm K[g(y)]\\\notag
&-2(a+y)\bm E[f(y,a)] -  2(a-y)\bm K[f(y,a)]\,,
\end{align}
with $g(y)=4y/(1+y)^2$ and $f(y,a)= 4ay/(a+y)^2$. 

The exchange energy gives rise to the following Coulomb-integral: 
\begin{align}
\notag
I_2(a)&=\int_a^1ydyI_1(a,y)=\frac {4}{3(1+a)}\Big[
2(1+a)(1+a^3)\\&-(1+a)^2(1+a^2)\bm E[h(a)]
+(1-a^2)^2\bm K[h(a)]\,
\Big]
\end{align}
with $h(a)= 4a/(1+a)^2$.

Both integrals involve the elliptic integrals $\bm E(m)$ and $\bm K(m)$
defined as
\begin{equation}
\bm E(m) = \int_0^{\pi/2}(1-m\sin^2\theta)^{1/2}d\theta\,,
\end{equation}
and
\begin{equation}
\bm K(m) = \int_0^{\pi/2}(1-m\sin^2\theta)^{-1/2}d\theta\,.
\end{equation}
\section{Analytical expression for the susceptibility}
\label{app_susc}
We will here present the approximate analytical result of the imaginary part
for the polarization. For this we define the following abbreviations:
$\tq=q/\sw$, $\tb=b/\sw$, $b_s=\tb+\sqrt{\tb^2+1}$ and
$b_c=\tb+\sqrt{\tb^2-1}$ using $\sinh^{-1}x=\ln(x+\sqrt{x^2+1})$ and
$\cosh^{-1}x=\ln(x+\sqrt{x^2-1})$. 
\subsection{Forward scattering}
For the analytic representation of the polarization of forward-scattering
processes, we define $f(x)=\cosh^{-1}(\tq x)$ and abbreviate $P=-2\pi
q\text{Im}P^{(1),\text{ret}}(\q,\omega)$.\\ 
For $0<q<b$, we then have
\begin{align}
\notag
P=\left\{\begin{array}{l l}
0&;\;\omega>q(2b+q)\\
f(b_s)\;&;\;q(2b-q)<\omega<q(2b+q)\\
f(b_s)-f(b_c)\;&;\;0<\omega<q(2b-q)
\end{array}\right.
\end{align}
For $b<q<2b$, we have
\begin{align}
\notag
P=\left\{\begin{array}{l l}
0&;\;\omega>q(2b+q)\\
f(b_s)\;&;\;b^2<\omega<q(2b+q)\\
f(b_s)-f(b_c)+f_F(b_c^{-1})\;&;\;q(2b-q)<\omega<b^2\\
f(b_s)-f(b_c)\;&;\;0<\omega<q(2b-q)
\end{array}\right.
\end{align}
\iffalse
For $2b<q<(1+\sqrt{2})b$, we have
\begin{align}
\text{Im}P^{(1),\text{ret}}(\q,\omega)=\frac{1}{2\pi}\left\{\begin{array}{l l}
0&;\;\omega>q(2b+q)\\
f_F(b_s)\;&;\;b^2<\omega<q(2b+q)\\
f_F(b_s)-f_F(b_c)+f_F(b_c^{-1})\;&;\;q(q-2b)<\omega<b^2\\
f_F(b_s)-f_F(b_s^{-1})-f_F(b_c)+f_F(b_c^{-1})\;&;\;0<\omega<q(q-2b)
\end{array}\right.
\end{align}
For $q>(1+\sqrt{2})b$, we have
\begin{align}
\text{Im}P^{(1),\text{ret}}(\q,\omega)=\frac{1}{2\pi}\left\{\begin{array}{l l}
0&;\;\omega>q(2b+q)\\
f_F(b_s)\;&;\;q(q-2b)<\omega<q(2b+q)\\
f_F(b_s)-f_F(b_s^{-1})\;&;\;b^2<\omega<q(q-2b)\\
f_F(b_s)-f_F(b_s^{-1})-f_F(b_c)+f_F(b_c^{-1})\;&;\;0<\omega<b^2
\end{array}\right.
\end{align}
\fi
For $q>2b$, we have:
\begin{align}
P&=
f(b_s)\Theta(q(q+2b)-\omega)-f(b_s^{-1})\Theta(q(q-2b)-\omega)\notag\\
&-\left[f(b_c)-f(b_c^{-1})\right]\Theta(b^2-\omega)
\notag
\end{align}
\subsection{Backward scattering}
For ``exact'' back-scattering $\q=2\kNull$, we have with $P=-2\pi
\omega^{1/4}\text{Im}P^{(1),\text{ret}}(2\kNull,\omega)$ the following
expression: 
\begin{align}
\notag
P=\sqrt{b_s}-\sqrt{b_s^{-1}}-\left[\sqrt{b_s}-\sqrt{b_s^{-1}}\right]
\end{align}

For ``positive'' back-scattering, i.e., $\q=2\kNull^+$ with
$|2\kNull^+|-2=q>0$, we define $f(x)=\cos^{-1}(\sqrt{\tq/x})$ and abbreviate
$P=-2\pi q^{1/2}\text{Im}P^{(1),\text{ret}}(2\kNull^+,\omega)$. 

For $q<b$, we have:
\begin{align}
P&=
f(b_s)-f(b_s^{-1})\Theta(\omega-q(2b+q))\notag\\
&-\left[f(b_c)-f(b_c^{-1})\Theta(\omega-q(2b-q)\right]\Theta(b^2-\omega)
\notag
\end{align}
\iffalse
For $0<q<(\sqrt{2}-1)b$, we have:
\begin{align}
\notag
P=\left\{\begin{array}{l l}
f_B^-(b_s)-f_B^-(b_s^{-1})&;\;\omega>b^2\\
f_B^-(b_s)-f_B^-(b_s^{-1})-f_B^-(b_c)+f_B^-(1)&;\;q(2b+q)<\omega<b^2\\
f_B^-(b_s)-f_B^-(b_c)+f_B^-(1)&;\;q(2b-q)<\omega<q(2b+q)\\
f_B^-(b_s)-f_B^-(b_c)&;\;0<\omega<q(2b-q)
\end{array}\right.
\end{align}
For $(\sqrt{2}-1)b<q<b$, we have:
\begin{align}
\notag
P=\left\{\begin{array}{l l}
f_B^-(b_s)-f_B^-(b_s^{-1})&;\;\omega>q(2b+q)\\
f_B^-(b_s)&;\;b^2<\omega<q(2b+q)\\
f_B^-(b_s)-f_B^-(b_c)+f_B^-(1)&;\;q(2b-q)<\omega<b^2\\
f_B^-(b_s)-f_B^-(b_c)&;\;0<\omega<q(2b-q)
\end{array}\right.
\end{align}
\fi
For $b<q<2b$, we have:
\begin{align}
\notag
P=\left\{\begin{array}{l l}
f(b_s)-f(b_s^{-1})&;\;\omega>q(2b+q)\\
f(b_s)&;\;q(2b-q)<\omega<q(2b+q)\\
f(b_s)-f(b_c)&;\;0<\omega<q(2b-q)
\end{array}\right.
\end{align}
For $q>2b$, we have:
\begin{align}
\notag
P=\left\{\begin{array}{l l}
f(b_s)-f(b_s^{-1})&;\;\omega>q(q+2b)\\
f(b_s)&;\;q(q-2b)<\omega<q(q+2b)\\
0&;\;0<\omega<q(q-2b)
\end{array}\right.
\end{align}\\
For ``negative'' back-scattering, i.e., $\q=2\kNull^-$ with
$2-|2\kNull^-|=q>0$, we define $f_s(x)=\sinh^{-1}(\sqrt{\tq x})$ and
$f_c(x)=\cosh^{-1}(\sqrt{\tq x})$. The polarization is given by
$P^1+P^2=-2\pi q^{1/2}\text{Im}P^{(1),\text{ret}}(2\kNull^-)$. The first part
is independent of the relative value of $q$ and $\omega$ and reads 
\begin{align}
\notag
P^1&=f_s(b_s)-f_s(b_s^{-1})
-\left[f_s(b_c)-f_s(b_c^{-1})\right]\Theta(b^2-\omega)\;.
\end{align}
For $0<q<b$, we further have:
\begin{align}
\notag
P^2=\left\{\begin{array}{l l}
0&;\;\omega>q(2b+q)\\
f_c(b_s)&;\;q(2b-q)<\omega<q(2b+q)\\
f_c(b_s)-f_c(b_c)&;\;0<\omega<q(2b-q)
\end{array}\right.
\end{align}
For $a<q<2b$, we have:
\begin{align}
\notag
P^2=\left\{\begin{array}{l l}
0&;\;\omega>q(2b+q)\\
f_c(b_s)&;\;b^2<\omega<q(2b+q)\\
f_c(b_s)-f_c(b_c)+f_c(b_c^{-1})&;\;q(2b-q)<\omega<b^2\\
f_c(b_s)-f_c(b_c)&;\;0<\omega<q(2b-q)
\end{array}\right.
\end{align}
For $q>2b$, we have:
\begin{align}
P^2&=
f_c(b_s)\Theta(q(q+2b)-\omega)-f_c(b_s^{-1})\Theta(q(q-2b)-\omega)\notag\\
&-\left[f_F(b_c)-f_F(b_c^{-1})\right]\Theta(b^2-\omega)
\notag
\end{align}
\iffalse
For $2b<q<(1+\sqrt{2})b$, we have:
\begin{align}
\text{Im}P^{(1),\text{ret},2}(2-q,\omega)=\frac{(\omega)^{1/4}}{4\pi}\left\{\begin{array}{l l}
0&;\;\omega>q(2b+q)\\
f_B^0(\tq^{-1})-f_B^0(b_s)&;\;b^2<\omega<q(2b+q)\\
f_B^0(\tq^{-1})-f_B^0(b_s)+g_B^0(b_c)+g_B^0(b_c^{-1})-2g_B^0(1)&;\;q(q-2b)<\omega<b^2\\
f_B^0(b_s^{-1})-f_B^0(b_s)+g_B^0(b_c)+g_B^0(b_c^{-1})-2g_B^0(1)&;\;0<\omega<q(q-2b)
\end{array}\right.
\end{align}
For $q>(1+\sqrt{2})b$, we have:
\begin{align}
\text{Im}P^{(1),\text{ret},2}(2-q,\omega)=\frac{(\omega)^{1/4}}{4\pi}\left\{\begin{array}{l l}
0&;\;\omega>q(2a+q)\\
f_B^0(\tq^{-1})-f_B^0(b_s)&;\;q(q-2b)<\omega<q(2b+q)\\
f_B^0(b_s^{-1})-f_B^0(b_s)&;\;b^2<\omega<q(q-2b)\\
f_B^0(b_s^{-1})-f_B^0(b_s)+g_B^0(b_c)+g_B^0(b_c^{-1})-2g_B^0(1)&;\;0<\omega<b^2
\end{array}\right.
\end{align}
\fi
%%%%%%%%%%%%%%%%%%%%%%%%%%%%%%%%%%%%%%%%%%%%%%%%%%%%%%%%%%%%%%%%%%%
% Section : bibliography
%%%%%%%%%%%%%%%%%%%%%%%%%%%

%%%%%%%%%%%%%%%%%%%%%%%%%%%%%%%%%%%%%%%%%%%%%%%%%%%%%%%%%%%%%%%%%%%
% Section : End of the document
%%%%%%%%%%%%%%%%%%%%%%%%%%%

\begin{thebibliography}{99}
\bibitem{note}Strictly speaking, marginal Fermi liquid behavior, as defined in the context of
  high temperature superconductors \cite{varma}, applies to systems with
  parabolic dispersion and a continuous Fermi surface. In that case, quasiparticles are not well defined
  because their spectral function is as broad as their energy.  Graphene, on the
  other hand, has linearly dispersing electrons, that is, Dirac fermions, and
  a Fermi surface consisting of points at the edge of the Brillouin zone. Hence,
  the concept of a marginal Fermi liquid is dull in their context.
\end{thebibliography}
\end{document}